# Polarized Positrons at Jefferson Lab

J. Dumas[a,b], J. Grames[b], E. Voutier[a]

[a]*Laboratoire de Physique Subatomique et de Cosmologie*
*IN2P3/CNRS – Université Joseph Fourier - INP*
*53, rue des Martyrs, 38026 Grenoble Cedex, France*

[b]*Thomas Jefferson National Accelerator Facility*
*12000 Jefferson Avenue, Newport News, VA 23606, USA*

**Abstract.** A novel concept for producing polarized positrons at Jefferson Lab using the CEBAF polarized electron photo-injector is presented. This approach relies on the polarization transfer from low energy highly polarized electrons to positrons via bremsstrahlung followed by pair production in a conversion target. An experiment to test this concept is discussed on the basis of GEANT4 simulations. It is shown that this low energy approach, which benefits from recent advances in high current high polarization electron sources, can yield positron longitudinal polarization up to 40%.

**Keywords:** Positron source, polarization transfer, electron sources.

**PACS:** 29.25.Bx, 41.75.Ht, 23.20.Ra, 07.77.Ka

## INTRODUCTION

The high spin polarization of the electron beam at the Continuous Electron Beam Accelerator Facility (CEBAF) at Jefferson Laboratory (JLab) is a significant feature of this facility. More generally, the polarization of a beam is a key element for the completion of a physics program and the achievement of precision physics. In addition, the sensitivity of electromagnetic processes to the charge of a lepton beam is a fundamental property, revealing singular features of a system when comparing negative and positive lepton responses. In the JLab context, a *polarized positron beam* would offer a unique opportunity for the study of Generalized Parton Distributions [1] or the two-photon exchange mechanisms, and globally the deep scattering processes.

An efficient scheme for positron production, widely used in particle accelerators, relies on the creation of electron-positron pairs from high energy photons. A significant aspect of the process is the dependence on the polarization, in particular, the circular polarization of the photon transfers to the longitudinal polarization of the positron [2]. A concept to generate polarized positrons from undulator-generated polarized photons [3] has recently been demonstrated by the E-166 experiment [4], in view of its application to the International Linear Collider (ILC) project. There, circularly polarized photons (~10 MeV) are created via synchrotron radiation using very high energy electrons (~50 GeV) travelling through a helical undulator, and are then converted into positrons via pair-creation. In contrast with this scheme, a new concept using a low energy (10 MeV) highly polarized electron beam to produce polarized positrons for the JLab physics program is hereafter discussed.

# JLAB POLARIZED POSITRON SOURCE CONCEPT

Similarly to pair creation, the bremsstrahlung process is a polarization sensitive mechanism. This property has been widely used at un-polarized electron accelerators to produce linearly polarized photon beams. In addition to the intrinsic linear polarization (L), the photons have a circular component (C) when the incoming electron beam is polarized, such that the bremsstrahlung of polarized electrons most generally lead to elliptically polarized photons [2,5]. The polarization component ratio C/L depends on the electron and photon energies, and is proportional to the electron beam polarization.

The proposed JLab polarized positron source concept takes advantage of recent advances of high-current high-polarization electron sources using GaAs-based photocathodes that simultaneously have both high electron polarization (>80%) and high quantum efficiency (~1%). Recent advances in electron gun performance have resulted in improved photocathode lifetime and sustainable high average current (1 mA) in a configuration suitable for an accelerator has been demonstrated [6]. At the CEBAF photo-injector a high intensity and highly polarized continuous electron beam would interact in a thin tungsten target foil to produce polarized positrons via bremsstrahlung followed by pair-creation. A thin target is chosen to limit the emittance of the positrons exiting the target as well as limit the extent of radiation. In this respect, a low electron beam energy (~10 MeV) helps limit photo-neutron production. The emitted positrons would then be selected in a given energy range, suitable for subsequent acceleration by the CEBAF system [7]. An experiment at the CEBAF photo-injector is presently considered to test and characterize this concept. Some key aspects are summarized in Table 1 along with those of the E-166 experiment. Practical ingredients for testing this concept are the pre-existing electron photo-injector and relatively small foot-print and cost for a conversion target and diagnostics beam line. An actual source for the CEBAF accelerator would be more sophisticated and costly, yet continue to be based upon the pre-existing photo-injector.

**TABLE 1.** Key aspects of the JLab and E-166 polarized positron generation schemes.

|  | JLab | E-166 [4] |
|---|---|---|
| Electron beam energy | ~10 MeV | ~50 GeV |
| Electron beam polarization | 85 % | Un-polarized |
| Photon production | Bremsstrahlung | Synchrotron |
| Converter target | Tungsten foil | Tungsten foil |
| Positron polarization | 40% (simulated) | 80% (measured) |

# SIMULATIONS

To study the proposed concept, particle interactions in a material are simulated with GEANT4 [8]. In particular, we expect the production of particles to follow the bremsstrahlung and pair creation cross-sections which are respectively given in [9] and [10]. One consideration is the incident electron energy. The electromagnetic shower is larger for high energy electrons in a high Z (atomic number) medium and photons created with higher energy are more likely to produce positrons, however, the energy distribution of bremsstrahlung photons decrease as $1/E_\gamma$. Another consideration is the transfer of polarization in an electromagnetic shower, also implemented in

GEANT4 [11]. Using Stokes parameters [5], each particle is followed, step-by-step, including the polarization transfer for bremsstrahlung and pair creation calculated by Olsen and Maximon [2], as well as depolarization due to bremsstrahlung, ionization, Compton and Moller/Bhabha scattering. For these processes, the polarization transfer is determined [3] as a function of the particle kinematics. These kinematic quantities are also combined into two parameters that determine the Coulomb and screening corrections of the cross-section and polarization transfer coefficients. These corrections to the Born approximation have been evaluated for the ultra-relativistic and small angle limits, which are known to be inaccurate at small energy [12]. In particular for the JLab concept, where energy distribution of the photon is near the pair-creation threshold, it is legitimate to question the effects of the mass-less electron and Lorentz focusing approximations. This problem is presently under study in view of an improved implementation in GEANT4. In the context of these simulations, calculations are limited to an ultra-relativistic Coulomb correction [13] and to the complete screening approximation, which presently appears in GEANT4 as the prescription in the considered low energy range.

Finally, GEANT4 was used to simulate the energy, yield and polarization distributions for a CEBAF electron beam with total energy of E = 5 MeV, longitudinal polarization P = 85% and current I = 1 mA. The target used in the simulation is a 250 μm thick tungsten foil. A contour plot of the distribution is shown in Fig. 1 where the positron energy is normalized to incident electron beam energy. In this simulation, first moments of the distribution yield a positron polarization of 43% and energy of 1.5 MeV.

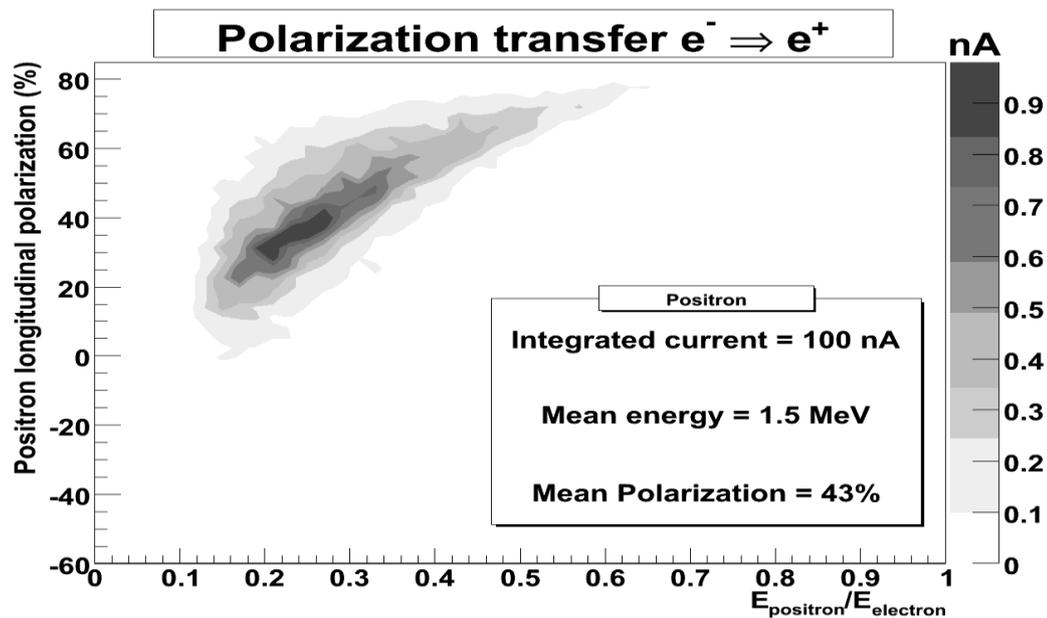

**FIGURE 1.** GEANT4 positron distribution of an electron beam E = 5 MeV, P = 85% and I = 1 mA incident on a 250 μm thick tungsten target foil.

# TEST OF CONCEPT

An experiment is planned to test the proposed concept, making use of the existing CEBAF electron photo-injector. A highly polarized electron beam (>80%) is produced using a strained superlattice GaAs/GaAsP photocathode inside a 100 kV DC high voltage electron gun. Electron bunches are emitted from the photocathode using a high-power fiber-based laser pulsing synchronously (1497 MHz) with the accelerating frequency. The low energy (100 keV) electrons are bunched, captured and accelerated to intermediate energy (<10 MeV) with small emittance ($\varepsilon_{n,rms}$ < 1 mm-mrad) and low energy spread ($\delta E/E < 2 \times 10^{-5}$). At this point the electron beam may be deflected to a Mott polarimeter to measure the electron polarization, a diagnostics beam line to measure the electron beam momentum (energy) or transverse profile, or toward the main accelerator. To test the polarized positron concept a new beam line is required containing the conversion target and means to characterize the positron distribution, limit the background and dump the primary electron beam.

The converter target is an important part of the positron source. The target should have a high atomic number (Z) to be more efficient for the production of an electromagnetic shower, but have sufficiently high thermal conductivity and melting point to dissipate deposited beam energy loss. Tungsten (Z = 74) is a good choice with a high melting point (3695 K) and can be made in a thin foil. A thinner target reduces positron yield, however, collection becomes easier because the electromagnetic shower is limited. As an example, the total integrated positron yield and deposited power versus target thickness resulting from an incident 5 MeV electron beam with current of 1 mA is shown in Fig. 2. The optimum positron yield ($Ne^+/Ne^-$ ~$10^{-4}$) occurs at a target thickness of 0.5 mm, corresponding to considerable power deposited in the target. To initially avoid high power issues a low duty factor pulsed electron beam may be used, however, optimization of target thickness, target cooling, positron yield and polarization are ultimately to be considered in any practical source operating a high average current.

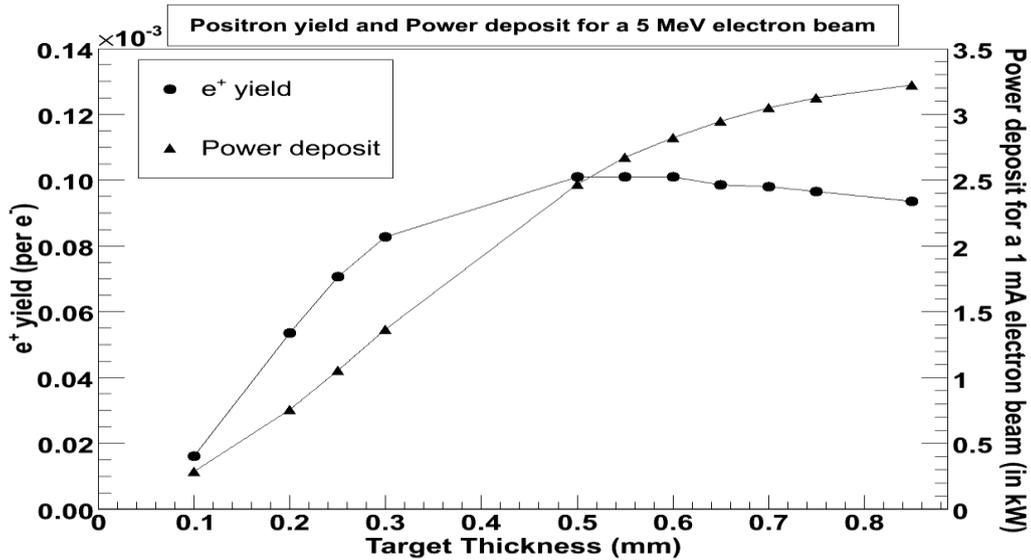

**FIGURE 2**: Positron yield (left axis) and deposited power (right axis) are plotted versus target thickness for a 5 MeV, 1 mA electron beam normally incident on a tungsten foil.

Ultimately, the shower of particles exiting the conversion target must be separated and isolated using a combination of spectrometer magnets and collimators to be analyzed and counted. Of interest are testing the positron and photon polarization when compared to the theory and simulation routines used by GEANT4, particularly at low energy and with respect to the aforementioned Coulomb and screening corrections. The intent is to measure the positron and photon polarization using a Compton transmission polarimeter [14].

## SUMMARY & OUTLOOK

A novel concept for producing polarized positrons at Jefferson Lab using the CEBAF polarized electron photo-injector is presented. The concept relies on transferring the electron beam spin polarization, via bremsstrahlung and pair production in a conversion target, to the positrons. Initial GEANT4 simulations indicate a positron polarization of 40% should be accessible with the present CEBAF photo-injector. An experiment to test this concept, benchmark modeling tools and collect valuable technical information for the design of an actual source is planned.